\documentclass[aps,prd,onecolumn,superscriptaddress,showpacs]{revtex4}

\usepackage{graphicx}
\usepackage{dcolumn}
\usepackage{bm}

\begin{document}

\newlength{\ogltemp}

\def\met{\settowidth{\ogltemp}{\big\slash}%
\hspace{0.5mm}\big\slash\hspace{-1.4\ogltemp}{E_T}%
}
\def\trkmet{\settowidth{\ogltemp}{\big\slash}%
\hspace{0.5mm}\big\slash\hspace{-1.4\ogltemp}{E_T^{trk}}%
}
\def\metvet{\settowidth{\ogltemp}{\big\slash}%
\hspace{0.5mm}\big\slash\hspace{-1.4\ogltemp}{\overrightarrow{E}_T}%
}
\def\trkmetvet{\settowidth{\ogltemp}{\big\slash}%
\hspace{0.5mm}\big\slash\hspace{-1.4\ogltemp}{\overrightarrow{E}_T^{trk}}%
}
\def\gev{\,\,\mathrm{GeV}\,}
\def\tev{\,\,\mathrm{TeV}\,}
\def\mev{\,\,\mathrm{MeV}\,}
\def\gevc{\,\,\mathrm{GeV}/c\,}
\def\gevm{\,\,\mathrm{GeV}/c^2\,}
\def\D0   {${\rm D}{\rm O\!\!\!\!\hspace{0.5pt}\protect\raisebox{0.2ex}{/}}$ }

\title{Search for Scalar Top Quark Production in $p\bar{p}$ Collisions at $\sqrt{s}=1.96$~TeV}

\affiliation{Institute of Physics, Academia Sinica, Taipei, Taiwan 11529, Republic of China}
\affiliation{Argonne National Laboratory, Argonne, Illinois 60439, USA}
\affiliation{University of Athens, 157 71 Athens, Greece}
\affiliation{Institut de Fisica d'Altes Energies, ICREA, Universitat Autonoma de Barcelona, E-08193, Bellaterra (Barcelona), Spain}
\affiliation{Baylor University, Waco, Texas 76798, USA}
\affiliation{Istituto Nazionale di Fisica Nucleare Bologna, $^{ee}$University of Bologna, I-40127 Bologna, Italy}
\affiliation{University of California, Davis, Davis, California 95616, USA}
\affiliation{University of California, Los Angeles, Los Angeles, California 90024, USA}
\affiliation{Instituto de Fisica de Cantabria, CSIC-University of Cantabria, 39005 Santander, Spain}
\affiliation{Carnegie Mellon University, Pittsburgh, Pennsylvania 15213, USA}
\affiliation{Enrico Fermi Institute, University of Chicago, Chicago, Illinois 60637, USA}
\affiliation{Comenius University, 842 48 Bratislava, Slovakia; Institute of Experimental Physics, 040 01 Kosice, Slovakia}
\affiliation{Joint Institute for Nuclear Research, RU-141980 Dubna, Russia}
\affiliation{Duke University, Durham, North Carolina 27708, USA}
\affiliation{Fermi National Accelerator Laboratory, Batavia, Illinois 60510, USA}
\affiliation{University of Florida, Gainesville, Florida 32611, USA}
\affiliation{Laboratori Nazionali di Frascati, Istituto Nazionale di Fisica Nucleare, I-00044 Frascati, Italy}
\affiliation{University of Geneva, CH-1211 Geneva 4, Switzerland}
\affiliation{Glasgow University, Glasgow G12 8QQ, United Kingdom}
\affiliation{Harvard University, Cambridge, Massachusetts 02138, USA}
\affiliation{Division of High Energy Physics, Department of Physics, University of Helsinki and Helsinki Institute of Physics, FIN-00014, Helsinki, Finland}
\affiliation{University of Illinois, Urbana, Illinois 61801, USA}
\affiliation{The Johns Hopkins University, Baltimore, Maryland 21218, USA}
\affiliation{Institut f\"{u}r Experimentelle Kernphysik, Karlsruhe Institute of Technology, D-76131 Karlsruhe, Germany}
\affiliation{Center for High Energy Physics: Kyungpook National University, Daegu 702-701, Korea; Seoul National University, Seoul 151-742, Korea; Sungkyunkwan University, Suwon 440-746, Korea; Korea Institute of Science and Technology Information, Daejeon 305-806, Korea; Chonnam National University, Gwangju 500-757, Korea; Chonbuk National University, Jeonju 561-756, Korea}
\affiliation{Ernest Orlando Lawrence Berkeley National Laboratory, Berkeley, California 94720, USA}
\affiliation{University of Liverpool, Liverpool L69 7ZE, United Kingdom}
\affiliation{University College London, London WC1E 6BT, United Kingdom}
\affiliation{Centro de Investigaciones Energeticas Medioambientales y Tecnologicas, E-28040 Madrid, Spain}
\affiliation{Massachusetts Institute of Technology, Cambridge, Massachusetts 02139, USA}
\affiliation{Institute of Particle Physics: McGill University, Montr\'{e}al, Qu\'{e}bec, Canada H3A~2T8; Simon Fraser University, Burnaby, British Columbia, Canada V5A~1S6; University of Toronto, Toronto, Ontario, Canada M5S~1A7; and TRIUMF, Vancouver, British Columbia, Canada V6T~2A3}
\affiliation{University of Michigan, Ann Arbor, Michigan 48109, USA}
\affiliation{Michigan State University, East Lansing, Michigan 48824, USA}
\affiliation{Institution for Theoretical and Experimental Physics, ITEP, Moscow 117259, Russia}
\affiliation{University of New Mexico, Albuquerque, New Mexico 87131, USA}
\affiliation{The Ohio State University, Columbus, Ohio 43210, USA}
\affiliation{Okayama University, Okayama 700-8530, Japan}
\affiliation{Osaka City University, Osaka 588, Japan}
\affiliation{University of Oxford, Oxford OX1 3RH, United Kingdom}
\affiliation{Istituto Nazionale di Fisica Nucleare, Sezione di Padova-Trento, $^{ff}$University of Padova, I-35131 Padova, Italy}
\affiliation{University of Pennsylvania, Philadelphia, Pennsylvania 19104, USA}
\affiliation{Istituto Nazionale di Fisica Nucleare Pisa, $^{gg}$University of Pisa, $^{hh}$University of Siena and $^{ii}$Scuola Normale Superiore, I-56127 Pisa, Italy}
\affiliation{University of Pittsburgh, Pittsburgh, Pennsylvania 15260, USA}
\affiliation{Purdue University, West Lafayette, Indiana 47907, USA}
\affiliation{University of Rochester, Rochester, New York 14627, USA}
\affiliation{The Rockefeller University, New York, New York 10065, USA}
\affiliation{Istituto Nazionale di Fisica Nucleare, Sezione di Roma 1, $^{jj}$Sapienza Universit\`{a} di Roma, I-00185 Roma, Italy}
\affiliation{Rutgers University, Piscataway, New Jersey 08855, USA}
\affiliation{Texas A\&M University, College Station, Texas 77843, USA}
\affiliation{Istituto Nazionale di Fisica Nucleare Trieste/Udine, I-34100 Trieste, $^{kk}$University of Udine, I-33100 Udine, Italy}
\affiliation{University of Tsukuba, Tsukuba, Ibaraki 305, Japan}
\affiliation{Tufts University, Medford, Massachusetts 02155, USA}
\affiliation{University of Virginia, Charlottesville, Virginia 22906, USA}
\affiliation{Waseda University, Tokyo 169, Japan}
\affiliation{Wayne State University, Detroit, Michigan 48201, USA}
\affiliation{University of Wisconsin, Madison, Wisconsin 53706, USA}
\affiliation{Yale University, New Haven, Connecticut 06520, USA}

\author{T.~Aaltonen}
\affiliation{Division of High Energy Physics, Department of Physics, University of Helsinki and Helsinki Institute of Physics, FIN-00014, Helsinki, Finland}
\author{B.~\'{A}lvarez~Gonz\'{a}lez$^z$}
\affiliation{Instituto de Fisica de Cantabria, CSIC-University of Cantabria, 39005 Santander, Spain}
\author{S.~Amerio}
\affiliation{Istituto Nazionale di Fisica Nucleare, Sezione di Padova-Trento, $^{ff}$University of Padova, I-35131 Padova, Italy}
\author{D.~Amidei}
\affiliation{University of Michigan, Ann Arbor, Michigan 48109, USA}
\author{A.~Anastassov$^x$}
\affiliation{Fermi National Accelerator Laboratory, Batavia, Illinois 60510, USA}
\author{A.~Annovi}
\affiliation{Laboratori Nazionali di Frascati, Istituto Nazionale di Fisica Nucleare, I-00044 Frascati, Italy}
\author{J.~Antos}
\affiliation{Comenius University, 842 48 Bratislava, Slovakia; Institute of Experimental Physics, 040 01 Kosice, Slovakia}
\author{G.~Apollinari}
\affiliation{Fermi National Accelerator Laboratory, Batavia, Illinois 60510, USA}
\author{J.A.~Appel}
\affiliation{Fermi National Accelerator Laboratory, Batavia, Illinois 60510, USA}
\author{T.~Arisawa}
\affiliation{Waseda University, Tokyo 169, Japan}
% ...

\collaboration{CDF Collaboration\footnote{With visitors from
$^a$Istituto Nazionale di Fisica Nucleare, Sezione di Cagliari, 09042 Monserrato (Cagliari), Italy,
$^b$University of CA Irvine, Irvine, CA 92697, USA,
$^c$University of CA Santa Barbara, Santa Barbara, CA 93106, USA,
$^d$University of CA Santa Cruz, Santa Cruz, CA 95064, USA,
$^e$Institute of Physics, Academy of Sciences of the Czech Republic, Czech Republic,
$^f$CERN, CH-1211 Geneva, Switzerland,
$^g$Cornell University, Ithaca, NY 14853, USA,
$^h$University of Cyprus, Nicosia CY-1678, Cyprus,
$^i$Office of Science, U.S. Department of Energy, Washington, DC 20585, USA,
$^j$University College Dublin, Dublin 4, Ireland,
$^k$ETH, 8092 Zurich, Switzerland,
$^l$University of Fukui, Fukui City, Fukui Prefecture, Japan 910-0017,
$^m$Universidad Iberoamericana, Mexico D.F., Mexico,
$^n$University of Iowa, Iowa City, IA 52242, USA,
$^o$Kinki University, Higashi-Osaka City, Japan 577-8502,
$^p$Kansas State University, Manhattan, KS 66506, USA,
$^q$Ewha Womans University, Seoul, 120-750, Korea,
$^r$University of Manchester, Manchester M13 9PL, United Kingdom,
$^s$Queen Mary, University of London, London, E1 4NS, United Kingdom,
$^t$University of Melbourne, Victoria 3010, Australia,
$^u$Muons, Inc., Batavia, IL 60510, USA,
$^v$Nagasaki Institute of Applied Science, Nagasaki, Japan,
$^w$National Research Nuclear University, Moscow, Russia,
$^x$Northwestern University, Evanston, IL 60208, USA,
$^y$University of Notre Dame, Notre Dame, IN 46556, USA,
$^z$Universidad de Oviedo, E-33007 Oviedo, Spain,
$^{aa}$CNRS-IN2P3, Paris, F-75205 France,
$^{bb}$Texas Tech University, Lubbock, TX 79609, USA,
$^{cc}$Universidad Tecnica Federico Santa Maria, 110v Valparaiso, Chile,
$^{dd}$Yarmouk University, Irbid 211-63, Jordan,
}}
\noaffiliation

\vfill\eject

\date{\today}

\begin{abstract}
We report on a search for the supersymmetric partner of the top quark
(scalar top) decaying into a charm quark and a neutralino in $p\bar{p}$ collisions 
at $\sqrt{s}=1.96$~TeV. The data sample, collected by the CDF~II
detector at the Fermilab Tevatron, corresponds to an integrated luminosity of $2.6$~fb$^{-1}$.
Candidate events are selected by requiring two or more 
jets and a large imbalance in the transverse momentum. To enhance the analysis sensitivity, 
at least one of the jets is required to be identified as originating
from a charm quark using an algorithm specifically designed for this analysis.
The selected events are in good agreement with standard model predictions. 
In the case of large mass splitting between the scalar top quark and the neutralino
we exclude a scalar top quark mass below $180\gevm$ at $95\%$~confidence level.
\end{abstract}


\maketitle

\section{Introduction}
The standard model (SM) of elementary particles and fundamental 
interactions, however successful, is still incomplete since it
does not explain the origin of electroweak symmetry breaking and
does not give an answer to the gauge hierarchy problem~\cite{Gildener:1976}.
A possible extension of the SM, supersymmetry
(SUSY)~\cite{Martin:1997ns}, solves these problems by introducing 
a symmetry that relates particles of different spin. 
R-parity~\cite{Martin:1997ns} conserving SUSY models also
provide a prime candidate for the dark matter in the
universe~\cite{Goldberg:1983}, namely the stable lightest
supersymmetric particle (LSP). In these models, the
left-handed and right-handed quarks have scalar partners,
respectively denoted as $\tilde{q}_L$ and $\tilde{q}_R$,
which can mix to form scalar
quarks with mass eigenstates $\tilde{q}_{1,2}$.
Several models~\cite{Bartl:1994bu} predict that this mixing can be
substantial for the scalar top (stop), yielding a stop
mass eigenstate~($\tilde{t}_{1}$) significantly lighter than other
scalar quarks. If the $\tilde{t}_{1}$ is sufficiently light, it can be 
pair-produced copiously in proton-antiproton collisions at center-of-mass energy
of $\sqrt{s}=1.96$ TeV at the Tevatron.
A stop in the mass range accessible at the
Tevatron is expected to decay, among other possibilities, into a charm quark
and the lightest neutralino ($\tilde{\chi}^{0}_{1}$), which is often
assumed to be the LSP. Assuming \mbox{R-parity} conservation, stops are produced in pairs. 
We consider a scenario where $m_{\tilde{t}_{1}} < m_{b} +
m_{\tilde{\chi}^{+}}$ (where $\tilde{\chi}^{+}$ is a chargino),  and $m_{\tilde{t}_{1}} < m_{W} + m_{b} +
m_{\tilde{\chi}^{0}_{1}}$. Under these conditions the decay
$\tilde{t}_{1} \rightarrow c\tilde{\chi}^{0}_{1}$ is dominant. 
Results from searches using this final state at the Tevatron have been 
previously reported in~\cite{stop_oldCDF, stop_D0}.

In this Letter, we report the search for $\tilde{t}_{1} \to c\tilde{\chi}^{0}_{1}$ 
decays in $p\bar{p}$ collision data from $2.6$~fb$^{-1}$ of integrated 
luminosity collected by the upgraded 
Collider Detector at Fermilab (CDF II) at the Tevatron.
The final state contains two $c$ jets from
the hadronization of the $c$ quarks and features 
an imbalance in transverse momentum (``missing transverse energy" or
$\met$~\cite{coordinates}) from the two LSPs escaping detection. 

\section{Analysis Description}
CDF II is a multipurpose detector, described in detail
elsewhere~\cite{CDFII}.
The charged-particle tracking system consists of silicon microstrip detectors
and a cylindrical open-cell drift chamber, both of which are immersed in a
$1.4$~T solenoidal magnetic field coaxial with the beam line.
The silicon detectors provide coverage in the 
pseudorapidity~\cite{coordinates} range $|\eta| \le 2$ and 
are used to identify events with long-lived particle decays. 
The drift chamber surrounds the silicon detectors, has maximum
tracking efficiency up to $|\eta| = 1$, and is used
for charged particle momentum measurements.
Segmented sampling calorimeters, which surround the tracking system, 
are arranged in a projective tower geometry and are used to measure
the energy of interacting particles in the region $|\eta| < 3.6$.
Muon candidates are identified by drift chambers, which extend up to $|\eta| = 1.5$, 
and are located outside the calorimeter volume.
Jets are reconstructed from the energy depositions in the
calorimeter cells using an iterative cone jet-clustering
algorithm~\cite{Arnison:1983rn}, with a cone size of
radius \mbox{$R = \sqrt{(\Delta\phi)^2+(\Delta \eta)^2} = 0.7$}~\cite{coordinates}.
Energy corrections~\cite{JES} are applied to account for effects such
as non-linear calorimeter response, underlying event, and the position
of the interaction point, that influence the measured transverse jet energy.

Candidate events used for this search are selected by an
online event selection system (trigger) that requires $\met\ge35\gev$
and two jets. Further selections are applied off-line to remove accelerator-produced and
detector-related backgrounds as well as cosmic-ray events.
After off-line event reconstruction, the events are required to have
$\met \ge 50$~GeV, and at least two jets with $|\eta| \le 2.4$ 
and $E_T \ge 25$~GeV. The highest-$E_T$ jet is required 
to have $E_T \ge 35$~GeV and at least one of the selected jets is required to 
be in the region $|\eta| \le 0.9$.

The hadrons in jets coming from $b$ or $c$ quark fragmentation, 
heavy-flavor (HF) jets, have a measurable flight path, 
yielding secondary vertices relative to the $p\bar{p}$ interaction point (primary vertex).
We require the events to have exactly one jet identified as a HF jet 
by the secondary-vertex tagging algorithm~\cite{Acosta:2004hw}, since
requiring two HF-identified jets enhances the sample with events
containing two $b$ quarks and reduces the signal acceptance. 
This criteria defines a preselection which is used as the basis of 
background studies and further optimization for the signal sample, as described below.


Dominant SM backgrounds are pair and single top-quark production, 
electroweak single and di-boson production, HF multijet production,
and light-flavor multijet events where one of the jets is falsely
tagged as a HF jet (mistag). 
The latter two background contributions are 
estimated from data. The \textsc{alpgen}~\cite{Alpgen} event generator interfaced
with the parton-shower model from the
\textsc{pythia}~\cite{Sjostrand:2000wi} event generator is used to
estimate the electroweak boson production, the
\textsc{madgraph}~\cite{madgraph} generator is used to model the
single top events, while the \textsc{pythia} event 
generator is used to model the top-quark pair and diboson backgrounds.
For the event generation the CTEQ5L~\cite{CTEQ5} 
parton distribution functions~(PDFs) are used. Simulated events are passed through the 
\textsc{geant3}-based~\cite{GEANT3} CDF II detector simulation~\cite{simulation} 
and are weighted by the probability that they would pass the trigger selection.
The single top-quark and diboson contributions are normalized 
to the theoretical cross sections~\cite{xsec_singtop,xsec_diboson1,xsec_diboson2}. 
The contributions for the electroweak boson samples 
are normalized to the next-to-leading order cross sections calculated
with \textsc{mcfm}~\cite{mcfm}. 
We use the measured top-quark pair production cross section 
of $\sigma_{t\bar{t}} = 7.02 \pm 0.63$~pb~\cite{top_xsec}. 
Mistags are estimated based on the mistag rate~\cite{Acosta:2004hw} 
observed in inclusive jet data. The mistag rate is parametrized as a function
of jet $E_{T}$, $|\eta|$, secondary-vertex track-multiplicity, the number of primary vertices in the event,
primary vertex $z$-position, and the scalar sum of $E_{T}$ of all jets in the event.
To estimate the HF multijet background from data, we use a multijet tag rate 
estimator (MUTARE) described elsewhere~\cite{gluino_sbottom,
  miguel_thesis}. The HF multijet prediction from MUTARE is scaled by a
multiplicative factor that is obtained in a signal-free region.

\begin{figure}

\includegraphics[width=7.5cm]{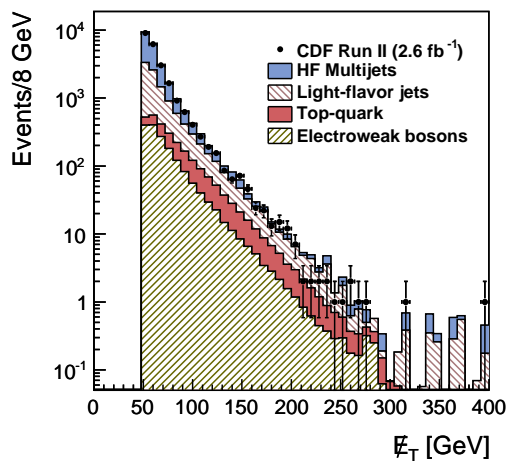}
\caption{ \label{met} Distribution of $\met$ in the pre-optimization region. SM prediction (stacked histograms) and observed distribution (dots) 
  are shown, where HF multijets and light-flavor jets are predicted from data.}
\end{figure}

To avoid potential biases when searching for new physics, we test
the various background contributions in distinct
control regions that are defined {\it a priori}. 
The three control regions used to validate the SM
prediction are denoted as multijet, lepton, and pre-optimization regions.
The multijet control region is defined to have the second leading
$E_T$ jet direction ($\vec{j}_{2}$) aligned with the $\metvet$~\cite{coordinates},
where aligned means $\Delta \phi (\metvet,\vec{j}_{2}) \leq 0.4$~rad.
This HF multijet enriched region is used to obtain the MUTARE parameterization to predict 
the HF multijet background in the other control and signal
regions. The lepton control region is defined to have $\vec{j}_{2}$
direction not aligned with the 
$\metvet$ ($\Delta \phi (\metvet,\vec{j}_{2}) \geq 0.7$~rad) and at
least one isolated charged lepton ($e$ or $\mu$)
with $p_{T} \geq 10 \gevc$. This lepton region
is used to validate the modeling of the top quark and electroweak $W/Z$ boson
backgrounds. The pre-optimization control region is 
defined to have the leading and second leading $E_T$ jets not aligned with the $\metvet$ 
and to have no identified leptons, where the lepton definition is the
same as in the lepton control region. We studied kinematic distributions
such as jet $E_T$, $\met$, as well as the total numbers of
events predicted. All these quantities are in agreement with
observations in the three control regions. As
an example, the $\met$ distributions for the pre-optimization region are shown 
in Fig.~\ref{met}. Table~\ref{tab:table1} presents the expected event yields for 
various background processes in the control regions under study.

\begin{table}
  \caption{\label{tab:table1} Comparison of the total number of expected
    and observed events in the control regions. The total uncertainty
    is computed by taking into account the (anti)correlations between
    the partial uncertainties. In the lepton region the HF multijet
    prediction is scaled to match the observed events
    in order to perform shape comparisons only.}
  \begin{center}
  \begin{tabular}{l c c c}
    \hline
  \hline
    Regions:                        & Multijet            &  Lepton        & Pre-optimization \\
    \hline

 $W/Z$ + jets        & $457 \pm 190$      &  $375 \pm 156$     & $1551 \pm 644$  \\
  Diboson              & $17 \pm 2$         &  $45 \pm 5$        & $118 \pm 13$  \\
  Top pair             & $188 \pm 21$       &  $547 \pm 60$      & $870 \pm 96$  \\
  Single top           & $11 \pm 2$         &  $71 \pm 10$       & $130 \pm 19$  \\
  HF multijets                     & $75407 \pm 23376$  &  $268 \pm 83$     & $12935 \pm 4010$  \\
  Light-flavor jets           & $65839 \pm 8427$   &  $720 \pm 92$      & $7741 \pm 991$  \\
  \hline
  Total expected                 & $141919 \pm 24849$ &  $2026 \pm 208$    & $23345 \pm 4182$  \\
  \hline
  Observed                       & 143441             &  2026              & 22792   \\
  \hline
  \hline
\end{tabular}
\end{center}
\end{table}

We optimize the sensitivity to stop production by applying additional
event selection criteria using a set of cuts
suppressing the HF multijet background.
We select events with exactly two jets, as expected from the signal, and 
fulfilling the condition $\Delta \phi(\metvet, \trkmetvet) < \pi/2$. 
This variable is the angular difference between the
calorimetry-based~$\metvet$ and the same quantity calculated using tracks
($\trkmetvet$) ~\cite{coordinates}. When the $\met$ in the event is real, these
two quantities are usually aligned in $\phi$. However, when the $\met$ 
comes from calorimetry mis-measurements, as in HF multijet events 
with no real $\met$, the angular difference between the two quantities is randomly distributed.

To further improve the sensitivity we apply a neural network (NN) trained with the \textsc{tmva}
package~\cite{TMVA}, to reduce the remaining HF multijet background.
We train the NN using jet $E_{T}$, 
jet $\eta$, the minimum $\Delta \phi$ between the $\metvet$ and
any of the selected jets,
$\Delta \phi (\metvet, \trkmetvet)$, $\Delta \phi (\vec{j}_{1}, \vec{j}_{2})$, 
$\met$, $\trkmet$, and the summed $E_T$ of all the jets in the event. 
We choose a reference signal point with  
$m(\tilde{t}_{1}) = 125 \gevm$ and $m(\tilde{\chi}^{0}_{1}) = 70 \gevm$
to perform the optimization. 
The signal acceptance is obtained using the \textsc{pythia} event
generator and CTEQ5L PDFs. Total signal yields are
normalized to the NLO production cross section determined with the \textsc{prospino} 
event generator~\cite{Beenakker:1996ch} and the
CTEQ6M~\cite{CTEQ6_1,CTEQ6_2} PDFs. 
The uncertainty of the NLO production cross section is estimated to be
20\%, arising from the scale dependence and the uncertainties on the
PDFs. The NN output lies within $-$1 and 1, where the background peaks at $-$1 and the
signal peaks at 1. We define our signal region as events with NN output scores~$>$~0.

The final stage in the optimization is the application of
a charm hadron analysis oriented separator (CHAOS) technique~\cite{miguel_thesis}, 
explicitly designed for this analysis to obtain a sample enriched in $c$~jets. 
CHAOS is a NN producing a two-dimensional output and trained with the \textsc{snns}~v4.3
package~\cite{SNNS} to determine whether a jet identified as HF has been
produced from the hadronization process of a light quark falsely tagged as
a HF jet, a $b$~quark, or a $c$~quark. The two-dimensional
output structure allows the separation of the three different targets 
during the same training process. Depending on the
flavor of the original parton, the jet identified as HF and its secondary vertex have 
different characteristics, mainly related to the tracking. Using
properties of the tracks forming the secondary vertex and the tracks
of the jets within a neural network, CHAOS allows enhancement of the jet
selection with a desired flavor, in particular $c$~jets.
We apply CHAOS to the jet identified as HF and 
we find that the optimal cut has a selection efficiency~\cite{miguel_thesis} of $34\%$ for
$c$~jets, $7.3\%$ for $b$~jets, and $4.9\%$ for light~jets.


The systematic uncertainties on the signal and the background
predictions, taking into account correlated and uncorrelated uncertainties, are studied.
Correlated uncertainties, affecting both the background prediction
and signal acceptance, are dominated by the uncertainties on the
performance of the $b$-tagging algorithm and CHAOS, which are
$4.4\%$~\cite{Acosta:2004hw} and $9.2\%$ respectively, 
and the luminosity~($6\%$)~\cite{CDFII}. Uncorrelated systematic uncertainties on the
background predictions are dominated by uncertainties on the MUTARE
parameterization ($30\%$), 
the mistag rate ($16\%$~\cite{Acosta:2004hw} for light-flavor multijets), 
the top-quark pair-production cross section ($11\%$), the single
top-quark production cross section ($13\%$), 
and the diboson production cross section ($10\%$ for $WW$/$WZ$ and
$20\%$ for $ZZ$). 
The uncertainty on the normalization of the boson plus HF jets to the
total inclusive boson production cross section translates into a $10\%$ uncertainty in the SM predictions.
Correlated and uncorrelated uncertainties are evaluated separately and combined in quadrature.

\section{Results}
The signal region is analyzed after the background predictions
are determined. We observe 115 events, where $132.0\pm 24.4$ are expected from
background, as summarized in Table~\ref{tab:table2} where yields 
before selecting events based on CHAOS output are also shown. 
Since no significant deviation from the SM prediction is observed, the results are used
to calculate a 95\%~C.L. exclusion limit for the $\tilde{t}_{1}$ pair production cross section.

\begin{figure}
\includegraphics[width=7.5cm]{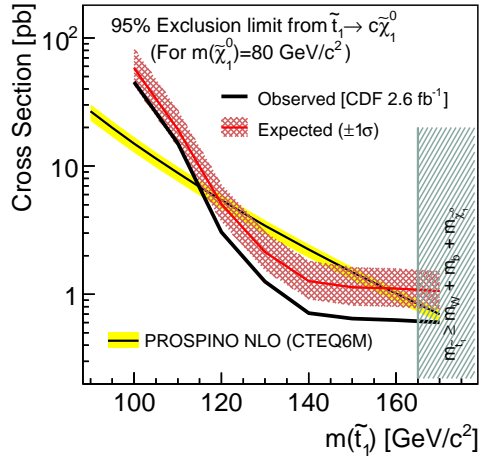}
\caption{Observed (solid line) and expected (solid line with dashed band) 95\% C.L. upper limit 
on the stop cross section (solid line with shaded band) as a function of
the stop mass for an assumed value of the neutralino mass. The shaded
band denotes the uncertainty on the NLO stop pair-production cross section.} 
\label{fig:limit_cs}
\end{figure}

We have used the differences in shape of the NN output to set the limits.
These limits are computed using a Bayesian likelihood method~\cite{limit} with a 
flat prior probability for the signal cross section and Gaussian priors for
the uncertainties on acceptance and backgrounds.
Figure~\ref{fig:limit_cs} shows the expected and observed limits as a
function of $m(\tilde{t}_{1})$ for a neutralino mass of $80\gevm$.


\begin{table}
\caption{\label{tab:table2} Number of expected and observed events
in the signal region before and after CHAOS application. Prediction for the signal
point ($m(\tilde{t}_{1}) = 125 \gevm$, $m(\tilde{\chi}^{0}_{1}) = 70
\gevm$) is also shown. Correlated and uncorrelated uncertainties in the total
background and expected signal were treated separately in the analysis 
although they are combined here.}
\begin{center}
\begin{tabular}{l c c}
  \hline
  \hline
   Signal region       & Before CHAOS   & Final (after CHAOS)   \\
  \hline
  $W/Z$ + jets        &  $423.6 \pm 185.0$  &  $ 60.9\pm 26.6$  \\
  Diboson                &  $36.9 \pm 6.5$        &  $ 10.7\pm 1.9$  \\
  Top pair                 &  $61.9 \pm 15.5$     &  $ 4.6\pm 1.3$\\
  Single top              &   $39.0 \pm 9.7$      &  $ 3.2\pm 0.8$ \\
  HF Multijets           &  $279.6 \pm 208.3$ &  $ 20.4\pm 15.2$\\
  Light-flavor jets  &  $658.3 \pm 259.6$  &  $ 32.2\pm 12.7$\\
  Total expected       &  $1499.3 \pm 277.1$ &  $ 132.0\pm 24.4$\\
  \hline 
  Observed                &  $1496$ &  $115$ \\
 \hline 
 $\tilde{t}_{1}$ signal & $250.0 \pm 66.2$ & $90.2 \pm 23.9$   \\
     \hline
     \hline
\end{tabular}
\end{center}
\end{table}


\begin{figure}
\includegraphics[width=7.5cm]{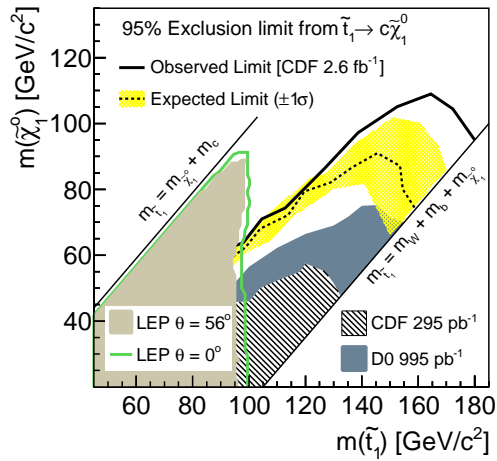}
\caption{Excluded region at 95\% C.L. in the
  $m(\tilde{t}_{1})$-$m(\tilde{{\chi}^{0}_{1}})$ plane. 
  The result is compared to the previous results from 
  CDF~\cite{stop_oldCDF}, from D0~\cite{stop_D0}, and from
  LEP~~\cite{stop_lep} experiments at CERN.}
\label{fig:limit_mass}
\end{figure}

We exclude, assuming BR$(\tilde{t}_{1} \to c\tilde{\chi}^{0}_{1}) = 100\%$, 
$\tilde{t}_{1}$ masses up to $180\gevm$ at 95\%~C.L. In addition, 
a 95\%~C.L. limit is obtained in the mass parameter plane of the model. 
Figure~\ref{fig:limit_mass} shows the excluded region in the stop-neutralino
mass plane of the analysis, compared with results from previous 
analyses~\cite{stop_oldCDF, stop_D0, stop_lep}.
The limit obtained with the present analysis improves the results of previous
searches using a similar topology, and represents the world's best limit in the region of large mass splitting.

\section{Summary}
To summarize, we have searched for the pair production of stop
decaying into a charm quark and a neutralino, in $2.6$~fb$^{-1}$ of
CDF Run II data.
We observe 115 candidate events, which are in
agreement with SM background expectations of $132.0 \pm 24.4$ events.
No evidence for stop is observed, and 
we exclude a region in the stop and neutralino
mass plane at $95\%$~C.L. as shown in Fig.~\ref{fig:limit_mass}.
Assuming BR$(\tilde{t}_{1} \to c\tilde{\chi}^{0}_{1}) = 100\%$, we 
exclude stop masses up to $180\gevm$ at $95\%$~C.L. for a neutralino
mass of  $90\gevm$.

\section*{Acknowledgements}
We thank the Fermilab staff and the technical staffs of the 
participating institutions for their vital contributions. 
This work was supported by the U.S. Department of Energy and 
National Science Foundation; the Italian Istituto Nazionale di 
Fisica Nucleare; the Ministry of Education, Culture, Sports, 
Science and Technology of Japan; the Natural Sciences and 
Engineering Research Council of Canada; the National Science 
Council of the Republic of China; the Swiss National Science 
Foundation; the A.P. Sloan Foundation; the Bundesministerium f\"ur 
Bildung und Forschung, Germany; the Korean World Class 
University Program, the National Research Foundation of Korea; 
the Science and Technology Facilities Council and the Royal Society, 
UK; the Russian Foundation for Basic Research; the Ministerio de 
Ciencia e Innovaci\'{o}n, and Programa Consolider-Ingenio 2010, 
Spain; the Slovak R\&D Agency; the Academy of Finland; and the 
Australian Research Council (ARC).

\end{document}